\documentclass[aps, pra, article, amsmath, amssymb, showpacs, superscriptaddress]{revtex4}

\usepackage{graphicx}
\usepackage{bm}

\begin{document}

\title{Generalized auto-balanced Ramsey spectroscopy of clock transitions}

\author{V. I. Yudin}
\email{viyudin@mail.ru}
\affiliation{Novosibirsk State University, ul. Pirogova 2, Novosibirsk, 630090, Russia}
\affiliation{Institute of Laser Physics SB RAS, pr. Akademika Lavrent'eva 13/3, Novosibirsk, 630090, Russia}
\affiliation{Novosibirsk State Technical University, pr. Karla Marksa 20, Novosibirsk, 630073, Russia}
\affiliation{National Institute of Standards and Technology, Boulder, Colorado 80305, USA}
\author{A. V. Taichenachev}
\affiliation{Novosibirsk State University, ul. Pirogova 2, Novosibirsk, 630090, Russia}
\affiliation{Institute of Laser Physics SB RAS, pr. Akademika Lavrent'eva 13/3, Novosibirsk, 630090, Russia}
\author{M.~Yu.~Basalaev}
\affiliation{Novosibirsk State University, ul. Pirogova 2, Novosibirsk, 630090, Russia}
\affiliation{Institute of Laser Physics SB RAS, pr. Akademika Lavrent'eva 13/3, Novosibirsk, 630090, Russia}
\affiliation{Novosibirsk State Technical University, pr. Karla Marksa 20, Novosibirsk, 630073, Russia}
\author{T.~Zanon-Willette}
\affiliation{LERMA, Observatoire de Paris, PSL Research University, CNRS, Sorbonne Universit$\acute{e}$s, UPMC Univ. Paris 06, F-75005, Paris, France}
\author{J.~W.~Pollock}
\affiliation{National Institute of Standards and Technology, Boulder, Colorado 80305, USA}
\author{M.~Shuker}
\affiliation{National Institute of Standards and Technology, Boulder, Colorado 80305, USA}
\author{E.~A.~Donley}
\affiliation{National Institute of Standards and Technology, Boulder, Colorado 80305, USA}
\author{J.~Kitching}
\affiliation{National Institute of Standards and Technology, Boulder, Colorado 80305, USA}

\date{\today}

\begin{abstract}
When performing precision measurements, the quantity being measured is often perturbed by the measurement process itself. This includes precision frequency measurements for atomic clock applications carried out with Ramsey spectroscopy. With the aim of eliminating probe-induced perturbations, a method of generalized auto-balanced Ramsey spectroscopy (GABRS) is presented and rigorously substantiated.
Here, the usual local oscillator frequency control loop is augmented with a second control loop derived from secondary Ramsey sequences interspersed with the primary sequences and with a different Ramsey period. This second loop feeds back to a secondary clock variable and ultimately compensates for the perturbation of the clock frequency caused by the measurements in the first loop. We show that such a two-loop scheme can lead to perfect compensation of measurement-induced light shifts and does not suffer from the effects of relaxation, time-dependent pulse fluctuations and phase-jump modulation errors that are typical of other hyper-Ramsey schemes. Several variants of GABRS are explored based on different secondary variables including added relative phase shifts between Ramsey pulses, external frequency-step compensation, and variable second-pulse duration. We demonstrate that a universal anti-symmetric error signal, and hence perfect compensation at finite modulation amplitude, is generated only if an additional frequency-step applied during both Ramsey pulses is used as the concomitant variable parameter. This universal technique can be applied to the fields of atomic clocks, high-resolution molecular spectroscopy, magnetically induced and two-photon probing schemes, Ramsey-type mass spectrometry, and to the field of precision measurements. Some variants of GABRS can also be applied for rf atomic clocks using CPT-based Ramsey spectroscopy of the two-photon dark resonance.

\end{abstract}


\maketitle

\section{Introduction}
Atomic clocks are based on high-precision spectroscopy of isolated quantum systems and are currently the most precise scientific instruments. Fractional frequency instabilities and accuracies at the level of 10$^{-18}$ have already been achieved, with the goal of 10$^{-19}$ on the horizon \cite{Schioppo_2017}. Frequency measurements at such a level could enable new tests of quantum electrodynamics and cosmological models, searches for drifts of fundamental constants, and new types of chronometric geodesy \cite{Ludlow_2015}.

For some of the promising clock systems, a key limitation is the frequency shift of the clock transition due to the excitation pulses themselves (probe-field-induced shift). In particular, for ultranarrow transitions (e.g., electric octupole \cite{hos09} and two-photon transitions \cite{fis04,badr06}), the off-resonant ac-Stark shift can be so large in some cases that high-accuracy clock performance is not possible. In the case of magnetically induced spectroscopy \cite{yudin06,bar06}, these shifts (quadratic Zeeman and ac-Stark shifts) could ultimately limit the achievable performance. A similar limitation exists for clocks based on direct frequency comb spectroscopy \cite{fortier06,stowe08} due to off-resonant ac-Stark shifts induced by large numbers of off-resonant laser modes. In addition to optical standards, probe-field-induced shifts can create significant instability for atomic clocks in the microwave range  based on coherent population trapping (CPT) \cite{Hemmer_JOSAB_1989, Shahriar_1997, Zanon_2005, Pati_2015, Hafiz_2017, Liu_2017}.

These challenges can be addressed through the use of Ramsey spectroscopy \cite{rams1950}, including its different generalizations and modifications. In contrast to continuous-wave spectroscopy, Ramsey spectroscopy has a large number of additional degrees of freedom connected with a wide assortment of parameters that can be precisely controlled: the durations of Ramsey pulses $\tau^{}_1$ and $\tau^{}_2$, the time of free evolution (dark time) $T$, the phase composition of Ramsey pulses (e.g., the use of composite pulses \cite{Levitt_1996}), a variety of Ramsey sequences (e.g., the use of three and more Ramsey pulses), different variants to build an error signal, etc.

Some modified Ramsey schemes for the suppression of the probe-field-induced shifts in atomic clocks were theoretically described in Ref.~\cite{yudin2010}, which proposed the use of pulses with different durations ($\tau^{}_1\neq\tau^{}_2$) and the use of composite pulses in place of the standard Ramsey sequence with two equal $\pi/2$-pulses. This ``hyper-Ramsey'' scheme has been successfully realised in an ion clock based on an octupole transition in Yb$^{+}$ (see Refs.~\cite{hunt12,huntemann2016}), where a suppression of the light shift by four orders of magnitude and an immunity against its fluctuations were demonstrated. Further developments of the hyper-Ramsey approach have used new phase variants to build error signals \cite{NPL_2015, Zanon_2014, Zanon_2016}. This has allowed for significant improvement in the efficiency of suppression of the probe-field-induced shifts in atomic clocks. However, as was shown in Ref.~\cite{Yudin_2016}, all previous hyper-Ramsey methods \cite{yudin2010,hunt12,huntemann2016,NPL_2015,Zanon_2016,Zanon_2015} are sensitive to  decoherence and spontaneous relaxation, which can  appreciably impede the achievement of  relative instability and inaccuracy at the level of $10^{-18}$  (or lower) in modern and future atomic clocks, for which the probe-field-induced shift is not negligible. To eliminate this disadvantage, a more complicated construction of the error signal was recently proposed in Ref.~\cite{Zanon_2017}, which requires four measurements for each frequency point (instead of two measurements for previous methods) with the use of different generalized hyper-Ramsey sequences presented in Ref.~\cite{Zanon_2016}. Nevertheless the method in Ref.~\cite{Zanon_2017} is not free from other disadvantages related to technical issues such as time dependent pulse area fluctuations and/or phase-jump modulation errors during the measurement of the error signal.

The above approaches \cite{yudin2010,hunt12,huntemann2016,NPL_2015,Zanon_2016,Zanon_2015,Zanon_2017} can be referred to as one-loop methods, because they use only one feedback loop and one error signal. However, frequency stabilization can also be realized with two feedback loops connected to Ramsey sequences with different dark periods $T_1$ and $T_2$ \cite{Yudin_2016,Morgenweg_2014,Sanner_2017}. For example, where a synthetic frequency protocol was proposed \cite{Yudin_2016}, which, in combination with the original hyper-Ramsey sequence \cite{yudin2010}, allows for substantial reduction in the sensitivity to decoherence and non-idealities of interrogation procedure. An alternative and effective approach called auto-balanced Ramsey spectroscopy was proposed and experimentally demonstrated in Ref.~\cite{Sanner_2017}, where in addition to the stabilization of the clock frequency $\omega$, a second loop feeding back on a variable phase during the second pulse was employed. Both of these two-loop methods \cite{Yudin_2016,Sanner_2017} strongly suppress probe-induced shifts of the measurement of the clock frequency.

In this paper, we present and rigorously substantiate a method of generalized auto-balanced Ramsey spectroscopy (GABRS), of which the intuitive approach realized in Ref.~\cite{Sanner_2017} is a particular case.
Our method uses a two-loop approach to feed back on and stabilizes the clock frequency $\omega$ as well as a second (concomitant) parameter $\xi$, which is an adjustable property of the first and/or second Ramsey pulses $\tau^{}_1$ and $\tau^{}_2$. To determine the error signals, it is necessary to use Ramsey sequences with two different dark times $T_1$ and $T_2$. The operation of GABRS consists of the correlated stabilisation of both variable parameters $\omega$ and $\xi$. In addition to the suppression of probe-field-induced shifts, the GABRS technique is protected against various processes of decoherence and also technical issues including time-dependent pulse area fluctuations (even more powerful than the common weak pulse area variation from previous schemes) and  phase-jump modulation errors needed to generate the error signal. This is in contrast to previous hyper-Ramsey schemes \cite{yudin2010,NPL_2015,Zanon_2016}, which can suffer from relaxation, time dependent pulse fluctuations, and phase-jump modulation errors. We consider several variants of GABRS with the use of different concomitant parameters $\xi$. It is found that the most optimal and universal variant is based on the frequency-step technique, when the concomitant parameter $\xi$ is equal to the varied additional frequency step $\Delta_{\rm step}$ during both Ramsey pulses $\tau^{}_1$ and $\tau^{}_2$. In this case, universal anti-symmetrical error signals are realized.

\section{General theory}

\begin{figure}[t]
\centerline{\scalebox{0.5}{\includegraphics{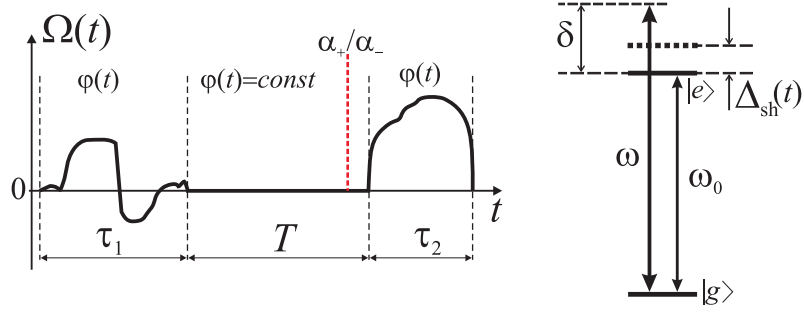}}}\caption{
Left part: schematic illustration of a sequence of two arbitrary Ramsey pulses (with durations $\tau^{}_1$ and $\tau^{}_2$) which are separated by the dark time $T$. Right part: scheme of the clock transition $|g\rangle\leftrightarrow |e\rangle$ (with unperturbed frequency $\omega_0$) interacting with the probe field at the frequency $\omega$.} \label{Fig1}
\end{figure}

In this section we demonstrate the universality and unprecedented robustness of GABRS. We will consider a two-level atom with unperturbed frequency $\omega_0$ of the clock transition $|g\rangle\leftrightarrow |e\rangle$ (see Fig.~\ref{Fig1}), which interacts with a Ramsey sequence of two absolutely arbitrary pulses (with durations $\tau^{}_1$ and $\tau^{}_2$) of the resonant probe field with frequency $\omega$:
\begin{equation}\label{E}
E(t)={\rm Re}\{{\cal E}(t)e^{-i\varphi (t)}e^{-i\omega t}\}\,,
\end{equation}
which are separated by a free evolution interval (dark time) $T$, during which the atom-field interaction is absent (see Fig.~\ref{Fig1}). We emphasise that the Ramsey pulses with arbitrary durations $\tau^{}_1$ and $\tau^{}_2$ can have an arbitrary shape and amplitude (i.e., during $\tau^{}_1$ and $\tau^{}_2$ an amplitude ${\cal E}(t)$ can be arbitrary real function), and an arbitrary phase function $\varphi(t)$ (e.g., the Ramsey pulses can be composite pulses). We assume only one restriction: aside from a phase modulation applied to generate the error signal (discussed below),  the phase function $\varphi (t)$ should be constant during the dark time $T$.

Our main goal consists of a development of a universal method, which allows us to stabilize the probe field frequency $\omega$ at the unperturbed frequency of the clock transition, $\omega=\omega_0$, in the presence of decoherence and arbitrary relaxation (including spontaneous). For this purpose, we will use the formalism of density matrix $\hat{\rho}$, which has the following form
\begin{equation}\label{rho_din}
\hat{\rho}(t)=\sum_{j,k=g,e}|j\rangle \rho_{jk}^{}(t)\langle k|\,,
\end{equation}
in the basis of states $|g\rangle$ and $|e\rangle$. In the resonance approximation, the density matrix components $\rho_{jk}^{}(t)$ satisfy the following differential equations:
\begin{eqnarray}\label{2_level}
&&[\partial^{}_t+\Gamma-i\tilde{\delta}(t)]\rho^{}_{eg}=i\Omega(t)[\rho^{}_{gg}-\rho^{}_{ee}]/2\,;\quad  \rho^{}_{ge}=\rho^{\ast}_{eg};\nonumber \\
&&[\partial^{}_t+\gamma^{}_{e} ] \rho^{}_{ee}-\gamma^{}_{g\to e} \rho^{}_{gg}=i[\Omega(t)\rho_{ge}-\rho_{eg}\Omega^{\ast}(t)]/2\,,\\
&&[\partial^{}_t+\gamma^{}_{g}]\rho^{}_{gg}-\gamma^{}_{e\to g} \rho^{}_{ee}=-i[\Omega(t)\rho_{ge}-\rho_{eg}\Omega^{\ast}(t)]/2\,.\nonumber
\end{eqnarray}

Here the time dependencies $\Omega(t)$ and $\tilde{\delta}(t)$ are
determined by the following: $\Omega(t)=\langle d\,\rangle {\cal E}(t)e^{-i\varphi (t)}$ and $\tilde{\delta}(t)=\delta-\Delta_{\rm sh}(t)$ during the action of Ramsey pulses $\tau^{}_1$ and $\tau^{}_2$, but $\Omega(t)=0$ and
$\tilde{\delta}(t)=\delta$ during the dark time $T$, $\langle d\,\rangle$ is a matrix element of the atomic dipole moment, $\delta=\omega-\omega_0$ is the detuning of the probe field from the unperturbed atomic frequency $\omega_0$, and $\Delta_{\rm sh}(t)$ is an actual probe-field-induced shift  (see Fig.~\ref{Fig1}) of the clock transition during the Ramsey pulses (e.g., it can be ac-Stark shift). Also Eq.~(\ref{2_level}) contains five relaxation constants, \{$\gamma^{}_{e}$, $\gamma^{}_{e\rightarrow g}$, $\gamma^{}_{g}$, $\gamma^{}_{g\rightarrow e}$, $\Gamma$\}: $\gamma^{}_{e}$ is a decay rate (e.g., spontaneous) of the exited state $|e\rangle$; $\gamma^{}_{e\rightarrow g}$ is a rate of the transmission (e.g., spontaneous) to the ground state $|g\rangle$; $\gamma^{}_{g}$ is a decay rate of the ground state $|g\rangle$ (e.g., due to black-body radiation and/or collisions); $\gamma^{}_{g\rightarrow e}$ is a rate of the transmission from the ground state $|g\rangle$ to the exited state $|e\rangle$. Note that $\gamma^{}_{e\rightarrow g}=\gamma^{}_{e}$ and $\gamma^{}_{g\rightarrow e}=\gamma^{}_{g}$ in the case of closed two-level system, while $\gamma^{}_{e\rightarrow g}<\gamma^{}_{e}$ and/or $\gamma^{}_{g\rightarrow e}<\gamma^{}_{g}$ in the case of open system. The constant $\Gamma=(\gamma^{}_{e}+\gamma^{}_{g})/2+\widetilde{\Gamma}$ describes the total rate of decoherence: spontaneous as well as all other processes, which are included in the parameter $\widetilde{\Gamma}$ (e.g., an influence of the nonzero spectral width of the probe field).

Equation~(\ref{2_level}) can be rewritten in the vector form:
\begin{equation}\label{vect_form}
\partial^{}_t \vec{\rho}(t)=\hat{L}(t)\vec{\rho}(t)\,,
\end{equation}
where $\hat{L}(t)$ is $4\times 4$ matrix, which is determined by the coefficients of Eq.~(\ref{2_level}), and $\vec{\rho}(t)$ is a vector formed by the matrix components $\rho_{jk}(t)$:
\begin{equation}\label{rho_vect}
\vec{\rho}(t)=\left(
             \begin{array}{c}
               \rho_{ee}(t) \\
               \rho_{eg}(t) \\
               \rho_{ge}(t) \\
               \rho_{gg}(t) \\
             \end{array}
           \right).
\end{equation}
In this case, a spectroscopic Ramsey signal can be presented in the following general form, which describes Ramsey fringes (as a function of $\delta$):
\begin{equation}\label{A_Rams}
A_{\rm Rams}(\delta)=(\vec{\rho}_{\rm obs}^{},\hat{W}^{}_{\tau^{}_2}\hat{G}^{}_T \hat{W}^{}_{\tau^{}_1}\vec{\rho}_{\rm in}^{})\,,
\end{equation}
where the scalar product is determined in the ordinary way: $(\vec{x},\vec{y})=\sum_{m}x^{*}_{m}y^{}_{m}$. Operators $\hat{W}^{}_{\tau^{}_1}$ and $\hat{W}^{}_{\tau^{}_2}$ describe an evolution of an atom during the first ($\tau^{}_1$) and second ($\tau^{}_2$) Ramsey pulses, respectively, and the operator $\hat{G}^{}_T$ describes free evolution during the dark time $T$. Vectors $\vec{\rho}_{\rm in}^{}$ and $\vec{\rho}_{\rm obs}^{}$ are initial and observed states, respectively. For example, if an atom before the Ramsey sequence was in the ground state $|g\rangle$, and after the Ramsey sequence we detect the atom in the exited state $|e\rangle$, then vectors $\vec{\rho}_{\rm in}^{}$ and $\vec{\rho}_{\rm obs}^{}$ are determined, in accordance with definition (\ref{rho_vect}), as the following:
\begin{equation}\label{in_obs}
\vec{\rho}_{\rm in}^{}=\left(
             \begin{array}{c}
               0 \\
               0 \\
               0 \\
               1 \\
             \end{array}
           \right),\quad
           \vec{\rho}_{\rm obs}^{}=\left(
             \begin{array}{c}
               1 \\
               0 \\
               0 \\
               0 \\
             \end{array}
           \right).
\end{equation}
However, for stabilization of the frequency $\omega$ we need to form an error signal (differential signal). In our approach, we use phase jumps $\alpha^{}_{+}$ and $\alpha^{}_{-}$ of the probe field before the second pulse $\tau^{}_2$ (see Fig.~\ref{Fig1}), as it was proposed in Ref.~\cite{mor89}. These jumps are described by the operators $\hat{\Phi}^{}_{+}$ and $\hat{\Phi}^{}_{-}$, respectively. As a result, the error signal can be presented as a difference:
\begin{equation}\label{err_gen}
S^{\rm (err)}_T = (\vec{\rho}_{\rm obs}^{},\hat{W}^{}_{\tau^{}_2}\hat{\Phi}^{}_{+}\hat{G}^{}_T \hat{W}^{}_{\tau^{}_1}\vec{\rho}_{\rm in}^{})-(\vec{\rho}_{\rm obs}^{},\hat{W}^{}_{\tau^{}_2}\hat{\Phi}^{}_{-}\hat{G}^{}_T \hat{W}^{}_{\tau^{}_1}\vec{\rho}_{\rm in}^{}) = (\vec{\rho}_{\rm obs}^{},\hat{W}^{}_{\tau^{}_2}\hat{D}^{}_{\Phi}\hat{G}^{}_T \hat{W}^{}_{\tau^{}_1}\vec{\rho}_{\rm in}^{})\,,
\end{equation}
with $\hat{D}^{}_{\Phi} = \hat{\Phi}^{}_{+}-\hat{\Phi}^{}_{-}$. To maximise the error signal, ${\alpha}_{\pm}=\pm \pi/2$ is typically used. However, in real experiments, we can have $|{\alpha}^{}_{+}|\neq |{\alpha}^{}_{-}|$ due to various technical reasons (e.g., electronics) which will lead to a shift of the stabilised frequency $\omega$ in the case of standard Ramsey spectroscopy. Therefore, here we will consider the general case of arbitrary ${\alpha}^{}_{+}$ and ${\alpha}^{}_{-}$ to demonstrate the robustness of  generalized auto-balanced Ramsey spectroscopy, where the condition $|{\alpha}^{}_{+}|\neq |{\alpha}^{}_{-}|$ will not lead to a frequency shift in atomic clocks.

Let us consider now the structure of the following operators: $\hat{G}^{}_T$, $\hat{\Phi}^{}_{+}$, $\hat{\Phi}^{}_{-}$, and $\hat{D}^{}_{\Phi}$. The operator of the free evolution $\hat{G}^{}_T$ has the following general matrix form:
\begin{equation}\label{GT_gen}
\hat{G}^{}_T=\left(
                 \begin{array}{cccc}
                   G_{11}(T) & 0 & 0 & G_{14}(T) \\
                   0 & e^{-(\Gamma -i\delta )T} & 0 & 0 \\
                   0 & 0 & e^{-(\Gamma +i\delta) T} & 0 \\
                   G_{41}(T) & 0 & 0 & G_{44}(T) \\
                 \end{array}
               \right),
\end{equation}
which corresponds to  Eq.~(\ref{2_level}) if $\Omega(t)=0$ and $\tilde{\delta}(t)=\delta$. The matrix elements $G_{11}(T)$, $G_{14}(T)$, $G_{41}(T)$, and $G_{44}(T)$ depend on four relaxation constants: \{$\gamma^{}_{e}$, $\gamma^{}_{e\rightarrow g}$, $\gamma^{}_{g}$, $\gamma^{}_{g\rightarrow e}$\}. In particular, for purely spontaneous relaxation of the exited state $|e\rangle$,  when $\gamma^{}_{g}=\gamma^{}_{g\rightarrow e}=0$, we obtain:
\begin{equation}\label{GT_sp}
\hat{G}^{}_T=\left(
                 \begin{array}{cccc}
                   e^{-\gamma^{}_{e}T} & 0 & 0 & 0 \\
                   0 & e^{-(\Gamma -i\delta )T} & 0 & 0 \\
                   0 & 0 & e^{-(\Gamma +i\delta) T} & 0 \\
                   \frac{\gamma^{}_{e\rightarrow g}}{\gamma^{}_{e}}(1-e^{-\gamma^{}_{e}T}) & 0 & 0 & 1 \\
                 \end{array}
               \right).
\end{equation}
Operators for the phase jumps $\hat{\Phi}^{}_{+}$ and $\hat{\Phi}^{}_{-}$ have forms:
\begin{equation}\label{Phi_pm}
\hat{\Phi}^{}_{\pm}=\left(
                 \begin{array}{cccc}
                   1 & 0 & 0 & 0 \\
                   0 & e^{i\alpha^{}_{\pm}} & 0 & 0 \\
                   0 & 0 & e^{-i\alpha^{}_{\pm}} & 0 \\
                   0 & 0 & 0 & 1 \\
                 \end{array}
               \right),
\end{equation}
which lead to the following expression for $\hat{D}^{}_{\Phi}$:
\begin{equation}\label{DPhi}
\hat{D}^{}_{\Phi}=\hat{\Phi}^{}_{+}-\hat{\Phi}^{}_{-}=\left(
                 \begin{array}{cccc}
                   0 & 0 & 0 & 0 \\
                   0 & (e^{i\alpha^{}_{+}}-e^{i\alpha^{}_{-}}) & 0 & 0 \\
                   0 & 0 & (e^{-i\alpha^{}_{+}}-e^{-i\alpha^{}_{-}}) & 0 \\
                   0 & 0 & 0 & 0 \\
                 \end{array}
               \right).
\end{equation}
As a result, taking into account Eq.~(\ref{GT_gen}), we obtain a formula for the matrix product $(\hat{D}^{}_{\Phi}\hat{G}^{}_T)$:
\begin{equation}\label{DV}
\hat{D}^{}_{\Phi}\hat{G}^{}_T=\left(
                 \begin{array}{cccc}
                   0 & 0 & 0 & 0 \\
                   0 & e^{-(\Gamma -i\delta )T}(e^{i\alpha^{}_{+}}-e^{i\alpha^{}_{-}}) & 0 & 0 \\
                   0 & 0 & e^{-(\Gamma +i\delta )T}(e^{-i\alpha^{}_{+}}-e^{-i\alpha^{}_{-}}) & 0 \\
                   0 & 0 & 0 & 0 \\
                 \end{array}
               \right)=e^{-\Gamma T}\hat{\Upsilon}^{}_{\delta T}\,,
\end{equation}
where the matrix $\hat{\Upsilon}^{}_{\delta T}$ is defined as
\begin{equation}
\hat{\Upsilon}^{}_{\delta T}=\left(
                 \begin{array}{cccc}
                   0 & 0 & 0 & 0 \\
                   0 & e^{i\delta T}(e^{i\alpha^{}_{+}}-e^{i\alpha^{}_{-}}) & 0 & 0 \\
                   0 & 0 & e^{-i\delta T}(e^{-i\alpha^{}_{+}}-e^{-i\alpha^{}_{-}}) & 0 \\
                   0 & 0 & 0 & 0 \\
                 \end{array}
               \right).
\end{equation}
Note that
\begin{equation}\label{Ups}
\hat{\Upsilon}^{}_{\delta T =0}=\hat{D}^{}_{\Phi}\,.
\end{equation}
Thus, the error signal (\ref{err_gen}) can be rewritten in the following form:
\begin{eqnarray}\label{err_new}
&& S^{\rm (err)}_T= e^{-\Gamma T}(\vec{\rho}_{\rm obs}^{},\hat{W}^{}_{\tau^{}_2}\hat{\Upsilon}^{}_{\delta T} \hat{W}^{}_{\tau^{}_1}\vec{\rho}_{\rm in}^{})\,.
\end{eqnarray}
Note that this result will be the same if we apply phase jumps $\alpha_{\pm}$ at any arbitrary point during the dark interval $T$.
It is interesting to note that the expression of the error signal in the presence of relaxation is formally different from the the error signal in the absence of relaxation only due to the scalar multiplier $e^{-\Gamma T}$, which affects the amplitude, first of all, but not the overall shape of the error signal. This is one of the main specific properties of the phase jump technique for Ramsey spectroscopy that makes it robust against relaxation. Indeed, for other well-known methods of frequency stabilisation, which use a {\em frequency} jump technique between alternating total periods of Ramsey interrogation $(\tau^{}_1+T+\tau^{}_2)$, relationship (\ref{err_gen}) does not exist. In addition, in the ideal case of $\alpha^{}_{+}=-\alpha^{}_{-}=\alpha$, the error signal (\ref{err_gen}) can be expressed as
\begin{eqnarray}\label{err_alpha}
&& S^{\rm (err)}_T=2\,{\rm sin}(\alpha) e^{-\Gamma T}(\vec{\rho}_{\rm obs}^{},\hat{W}^{}_{\tau^{}_2}\hat{\Theta}^{}_{\delta T} \hat{W}^{}_{\tau^{}_1}\vec{\rho}_{\rm in}^{})\,,
\end{eqnarray}
where the matrix $\hat{\Theta}^{}_{\delta T}$:
\begin{equation}
\hat{\Theta}^{}_{\delta T}=\left(
                 \begin{array}{cccc}
                   0 & 0 & 0 & 0 \\
                   0 & ie^{i\delta T} & 0 & 0 \\
                   0 & 0 & -ie^{-i\delta T} & 0 \\
                   0 & 0 & 0 & 0 \\
                 \end{array}
               \right),
\end{equation}
depends only on $\delta T$.

The main idea of GABRS is the following. First of all, apart from $\delta$ (i.e., frequency $\omega$) for the frequency stabilization procedure we will use some additional (concomitant) variable parameter $\xi$, which is related to the first and/or second Ramsey pulses $\tau^{}_1$ and $\tau^{}_2$. For example, the parameter $\xi$ can be equal to the phase $\phi^{c}$ of the second pulse as it was proposed in \cite{Sanner_2017}. However, as shown below, there are many other variants of the concomitant parameter $\xi$. Thus, the error signal in Eq. (\ref{err_new}) should be considered as a function of two variable parameters $S^{\rm (err)}_T(\delta,\xi)$. Secondly, we will use the Ramsey interrogation of the clock transition for two different, fixed intervals of free evolution $T_1$ and $T_2$, i.e., we will use two error signals $S^{\rm (err)}_{T_1}(\delta,\xi)$ and $S^{\rm (err)}_{T_2}(\delta,\xi)$.

For GABRS, the procedure for the frequency stabilization is organized as a series of the following cycles. For interrogation with dark time $T_1$, the parameter $\xi$ is fixed, and we stabilize the variable detuning $\delta$ (i.e., frequency $\omega$) at the zero point of the error signal: $S^{\rm (err)}_{T_1}(\delta,\xi_{fixed}^{})=0$. After this procedure, we switch to interrogation with dark time $T_2$, where we fix the previously obtained detuning $\delta$ and stabilize the variable parameter $\xi$ at the zero point of the second error signal: $S^{\rm (err)}_{T_2}(\delta_{fixed}^{},\xi)=0$. If we continue these cycles, then the final result (formally for $t\rightarrow \infty$) consists of the stabilization of both parameters, $\delta=\bar{\delta}_{\rm clock}$ and $\xi=\bar{\xi}$, which correspond to the solution of a system of two equations:
\begin{equation}\label{Sys_gen}
S^{\rm (err)}_{T_1}(\delta,\xi)=0\,,\quad S^{\rm (err)}_{T_2}(\delta,\xi)=0\,,
\end{equation}
in relation to the two unknowns $\delta$ and $\xi$. The value $\bar{\delta}_{\rm clock}$ describes the frequency shift in an atomic clock.

Taking into account  relationship (\ref{err_new}), the system Eq.~(\ref{Sys_gen}) can be written in the following form:
\begin{equation}\label{Sys_GABRS}
(\vec{\rho}_{\rm obs}^{},\hat{W}^{}_{\tau^{}_2}\hat{\Upsilon}^{}_{\delta T_1} \hat{W}^{}_{\tau^{}_1}\vec{\rho}_{\rm in}^{})=0\,,\;\;(\vec{\rho}_{\rm obs}^{},\hat{W}^{}_{\tau^{}_2}\hat{\Upsilon}^{}_{\delta T_2} \hat{W}^{}_{\tau^{}_1}\vec{\rho}_{\rm in}^{})=0\,.
\end{equation}
Let us show that Eq.~(\ref{Sys_GABRS}) always contains the solution $\delta=0$. Indeed, if we apply $\delta=0$ for operators $\hat{\Upsilon}^{}_{\delta T_1}$ and $\hat{\Upsilon}^{}_{\delta T_2}$, then due to Eq.~(\ref{Ups}) we obtain that the system of two equations (\ref{Sys_GABRS}) is reduced to the following single equation:
\begin{equation}\label{eq_alpha}
(\vec{\rho}_{\rm obs}^{},\hat{W}^{}_{\tau^{}_2}\hat{D}^{}_{\Phi} \hat{W}^{}_{\tau^{}_1}\vec{\rho}_{\rm in}^{})|^{}_{\delta=0}=0\,,
\end{equation}
in relation to only one unknown $\xi$, which always has a solution under appropriate choice of the parameter $\xi$.

Thus, we have analytically shown that the GABRS method always leads to zero field-induced shift of the stabilized frequency $\omega$ in an atomic clock, $\bar{\delta}_{\rm clock}=0$. This fundamental result does not depend on relaxation constants \{$\gamma^{}_{e}$, $\gamma^{}_{e\rightarrow g}$, $\gamma^{}_{g}$, $\gamma^{}_{g\rightarrow e}$, $\Gamma$\}, the values of phase jumps $\alpha^{}_{+}$ and $\alpha^{}_{-}$ used for error signals, or the parameters (such as: amplitude, shape, duration, phase structure $\varphi (t)$, shift $\Delta_{\rm sh}(t)$, etc.) of the two Ramsey pulses $\tau^{}_1$ and $\tau^{}_2$. Such a robustness is unprecedented for Ramsey spectroscopy. Indeed, all known methods of hyper-Ramsey spectroscopy \cite{yudin2010,hunt12,Zanon_2015,NPL_2015,Zanon_2016,Zanon_2017}, which can significantly suppress field-induced shifts, are sensitive (excepting Ref.~\cite{Zanon_2017}) to  relaxation processes and decoherence (see Ref.~\cite{Yudin_2016}), and all these methods require the use of rectangularly shaped Ramsey pulses. Moreover, all previously used Ramsey methods (including the usual Ramsey spectroscopy with two equal $\pi/2$-pulses) require the condition $\alpha^{}_{-}=-\alpha^{}_{+}$ for phase jumps, because any non-ideality ($\alpha^{}_{-}\neq -\alpha^{}_{+}$) will lead to an additional shift, which is approximately equal to the value of $-(\alpha^{}_{+} +\alpha^{}_{-})/(2T)$. Summarizing, practically all non-idealities of the interrogation procedure (including field-induced shifts of atomic levels) and relaxation processes (including decoherence) only influence  the stabilized concomitant parameter $\bar{\xi}$, while the stabilized frequency $\omega$ remains unshifted, with $\bar{\delta}_{\rm clock}=0$.

It is interesting to note that the solution of Eqs.~(\ref{Sys_GABRS})-(\ref{eq_alpha}) does not formally depend on the values $T_1$ and $T_2$ at all. However, from an experimental viewpoint it is better to use the condition $T_2\ll T_1$. Indeed, because during the interrogation procedure with dark time $T_1$ we stabilize the frequency $\omega$ using the error signal $S^{\rm (err)}_{T_1}(\delta,\xi_{fixed}^{})=0$,  we always have $|\delta |< 1/T_1$ even during the first cycles of the clock stabilization. On the other hand, nonzero detuning $\delta \neq 0$ will influence the second interrogation procedure with dark time $T_2$ (to stabilize the concomitant parameter $\xi$) in conformity with the value $\delta T_2$, which is contained in the error signal $S^{\rm (err)}_{T_2}(\delta,\xi)$. Therefore, if $T_2\ll T_1$, then we obtain an estimation: $|\delta T_2|< (T_2/T_1) \ll 1$, i.e., the results of the stabilization of the concomitant parameter $\xi$ (using $S^{\rm (err)}_{T_2}(\delta_{fixed}^{},\xi)=0$) will weakly depend on the results of the frequency stabilization during interrogation procedure with dark time $T_1$. An additional advantage of the condition $T_2\ll T_1$ is connected with the short-term stability of an atomic clock. Indeed, because the second feedback loop (stabilization of $\xi$) increases the total period of each cycle, then it is better to use shortest possible $T_2$. Formally we can even use $T_2=0$ (with the phase jumps $\alpha_{\pm}$ in the virtual point between pulses $\tau^{}_1$ and $\tau^{}_2$). However, due to technical transient regimes (i.e., in acousto-optic modulators) under switching off/on of Ramsey pulses in real experiments, we believe that it is necessary to keep some nonzero dark time, $T_2 \neq 0$, which significantly exceeds any various transient times. For example, in the case of magnetically-induced spectroscopy \cite{yudin06,bar06}, the transient processes, associated with switching off/on of magnetic field, can be relatively slow.

Though the solution $\bar{\delta}_{\rm clock}=0$ does not depend on the amplitude and shape of the Ramsey pulses, nevertheless, to maximize the error signals $S^{\rm (err)}_{T_1}(\delta,\xi_{fixed}^{})$ and $S^{\rm (err)}_{T_2}(\delta_{fixed}^{},\xi)$ we need to use quite specific types of the Ramsey pulses. Some appropriate variants are presented below.

\section{Different variants of GABRS}
In this section we consider some variants of Ramsey sequences with different choices for  the concomitant parameter $\xi$. Because arbitrary relaxation and practically all non-idealities of the Ramsey interrogation scheme do not lead to a shift of the stabilized clock frequency, $\bar{\delta}_{\rm clock}=0$, we will focus our attention only on the field-induced shift of the clock transition $\Delta_{\rm sh}$ during Ramsey pulses $\tau^{}_1$ and $\tau^{}_2$. We will show how the valu\emph{}e $\Delta_{\rm sh}$ influences the stabilized concomitant parameter $\bar{\xi}$ and  the error signals $S^{\rm (err)}_{T_1}(\delta,\xi=\bar{\xi})$ and $S^{\rm (err)}_{T_2}(\delta=0,\xi)$, which contain the main information about the dynamic efficiency of GABRS. For simplicity, all calculations are done for $\alpha_{\pm}=\pm\pi/2$ and in the absence of relaxation: $\gamma^{}_{e}=\gamma^{}_{e\rightarrow g}=\gamma^{}_{g}=\gamma^{}_{g\rightarrow e}=\Gamma=0$. The initial and observed states $\vec{\rho}_{\rm in}^{}$ and $\vec{\rho}_{\rm obs}^{}$  correspond to Eq.~(\ref{in_obs}).

\subsection{Auto-balanced Ramsey spectroscopy with additional phase correction}
\begin{figure}[t]
\centerline{\scalebox{0.5}{\includegraphics{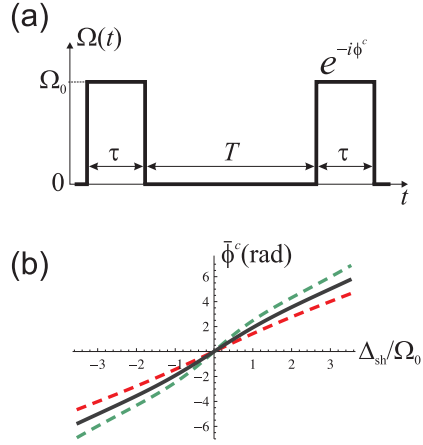}}}\caption{
(a) Schematic illustration of Ramsey pulses (with the same duration $\tau^{}$) for the auto-balanced Ramsey spectroscopy technique demonstrated in Ref.\cite{Sanner_2017}, where the concomitant parameter $\xi$ is equal to the additional phase $\phi^{c}$ of the second pulse.\\
(b) The dependencies of stabilized phase $\bar{\phi}^{c}(\Delta_{\rm sh})$ for different pulse area: $\Omega_0 \tau^{}=\pi/2$ (black solid line); $\Omega_0 \tau^{}=1.2\times\pi/2$ (green dashed line); $\Omega_0 \tau^{}=0.8\times\pi/2$ (red dashed line).} \label{Fig2}
\end{figure}

Here we describe a detailed theoretical basis for the original auto-balanced Ramsey spectroscopy method demonstrated in Ref.~\cite{Sanner_2017}. In the context of the general theory developed above, this spectroscopy can be considered as a partial case of GABRS, where the concomitant parameter $\xi$ is equal to the varied additional phase $\phi^{c}$ during the second pulse (Fig.~\ref{Fig2}(a)). In this case, we always have $\bar{\delta}_{\rm clock}=0$, and the stabilized phase $\bar{\phi}^{c}$ is determined as the solution of Eq.~(\ref{eq_alpha}). In the presence of the probe-field-induced shift of the clock transition $\Delta_{\rm sh}$ during the Ramsey pulses, the phase $\bar{\phi}^{c}$ is a function $\bar{\phi}^{c}(\Delta_{\rm sh})$ of the value $\Delta_{\rm sh}$. These dependencies are presented in Fig.~\ref{Fig2}(b) for different pulse areas $\Omega_0 \tau^{}$. In the case of $(\Delta_{\rm sh}/\Omega_0) <1$, we have the following approximate dependence: $\bar{\phi}^{c}(\Delta_{\rm sh})\approx 2r\Delta_{\rm sh}/\Omega_0$, where the coefficient $r$ determines the pulse area, $\Omega_0\tau^{} =r\pi/2$. Thus, this dependence can be written as: $\bar{\phi}^{c}(\Delta_{\rm sh})\approx 4\Delta_{\rm sh}\tau^{} /\pi$ (if $\Delta_{\rm sh}\tau^{} <1$).

\begin{figure}[t]
\centerline{\scalebox{0.5}{\includegraphics{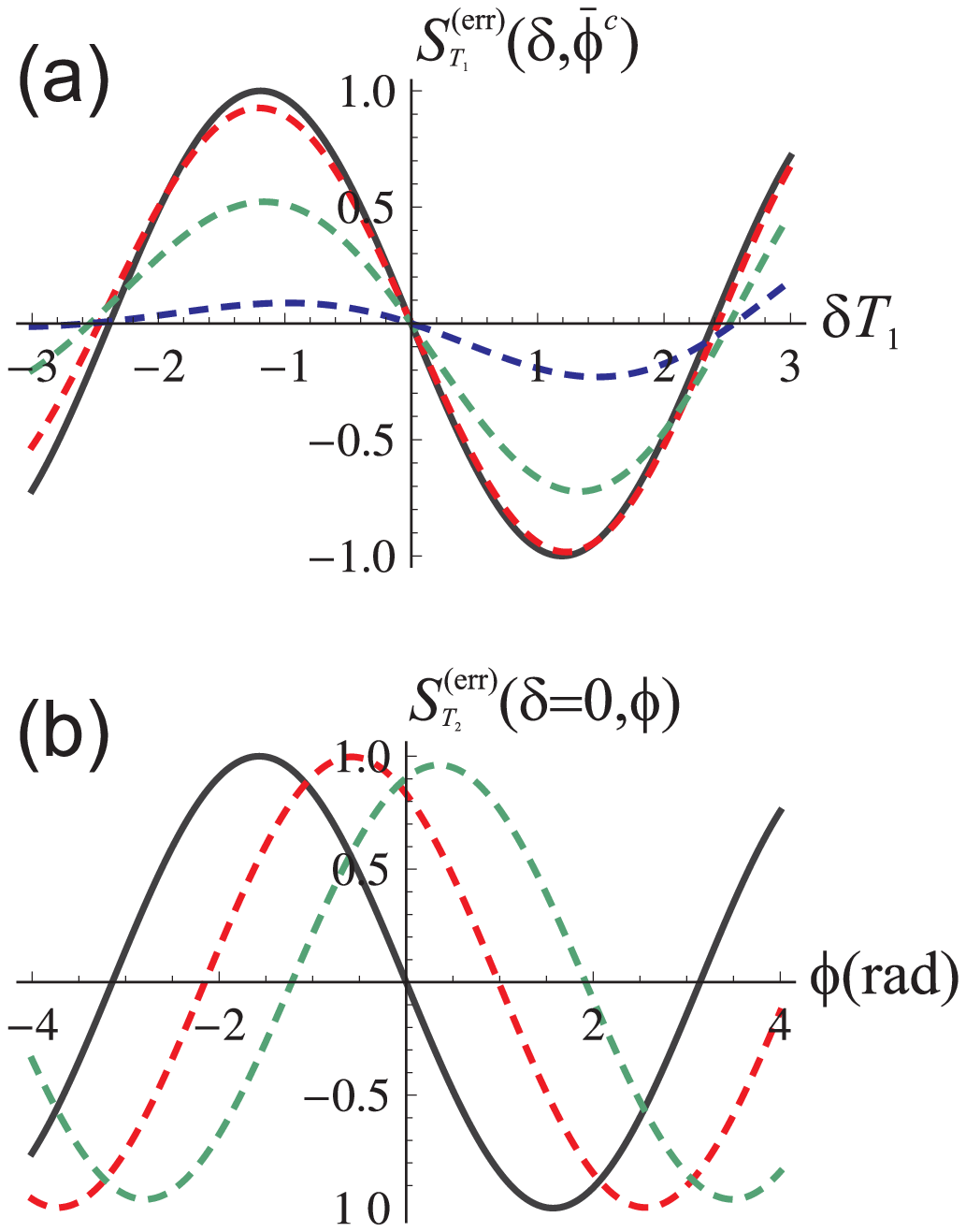}}}\caption{
Error signals under $\Omega_0 T_1=2\pi$, $\Omega_0 \tau^{}=\pi/2$, and for different field-induced shifts of the clock transition during Ramsey pulses, $\Delta_{\rm sh}$:\\
(a) Error signal $S^{\rm (err)}_{T_1}(\delta,\phi=\bar{\phi}^{c})$: $\Delta_{\rm sh}/\Omega_0=0$ (black solid line); $\Delta_{\rm sh}/\Omega_0=1$ (red dashed line); $\Delta_{\rm sh}/\Omega_0=2$ (green dashed line); $\Delta_{\rm sh}/\Omega_0=3$ (blue dashed line). \\
(b) Error signal $S^{\rm (err)}_{T_2}(\delta=0,\phi)$: $\Delta_{\rm sh}/\Omega_0=0$ (black solid line); $\Delta_{\rm sh}/\Omega_0=0.5$ (red dashed line); $\Delta_{\rm sh}/\Omega_0=1.0$ (green dashed line).} \label{Fig3}
\end{figure}

The error signals $S^{\rm (err)}_{T_1}(\delta,\phi=\bar{\phi}^{c})$ and $S^{\rm (err)}_{T_2}(\delta=0,\phi)$ for different values $\Delta_{\rm sh}$ are presented in Fig.~\ref{Fig3}. As we see, for the condition $|\Delta_{\rm sh}/\Omega_0 |>1$ the error signal $S^{\rm (err)}_{T_1}(\delta,\phi=\bar{\phi}^{c})$ becomes smaller and distinctly non-antisymmetrical, which can lead to clock errors. Thus, the auto-balancing technique of only varying the phase during the second Ramsey pulse works well only for $|\Delta_{\rm sh}/\Omega_0 | < 1$. Distortions in the error signals arising from this problem can be largely reduced by the use of an additional and well-controllable frequency step $\Delta_{\rm step}$ only during the Ramsey pulses $\tau^{}_1$ and $\tau^{}_2$ \cite{tai09,yudin2010}. In this case, all dependencies presented in Fig.~\ref{Fig2}(b) and \ref{Fig3} will be the same if we will replace $\Delta_{\rm sh}\rightarrow \Delta_{\rm eff}=(\Delta_{\rm sh}-\Delta_{\rm step})$.
Thus, we can always apply a frequency step $\Delta_{\rm step}$ (e.g., with an acousto-optic modulator) during excitation to achieve the condition $|\Delta_{\rm eff}/\Omega_0 | \ll 1$ for an effective shift $\Delta_{\rm eff}$, as it was used in experiments  \cite{hunt12,huntemann2016,NPL_2015,Sanner_2017}.

In addition, this variant of GABRS can also be  used in atomic clocks based on coherent population trapping (CPT), where we can use as the concomitant parameter $\xi$ the varied phase $\phi^{c}$ of the second (detecting) pulse in CPT-Ramsey spectroscopy.

\subsection{Auto-balanced Ramsey spectroscopy with additional frequency step}
\begin{figure}[t]
\centerline{\scalebox{0.5}{\includegraphics{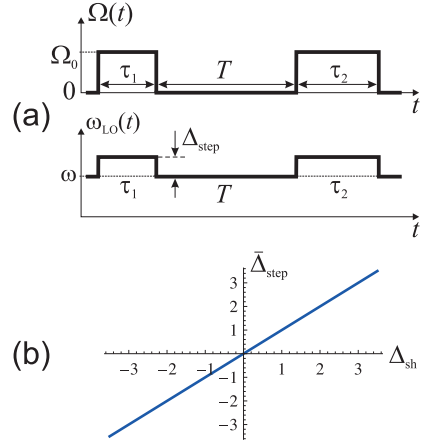}}}\caption{
(a) Schematic illustration of a Ramsey interrogation scheme for GABRS, where the concomitant parameter $\xi$ is equal to the additional frequency step $\Delta_{\rm step}$ during both Ramsey pulses $\tau^{}_1$ and $\tau^{}_2$.\\
(b) The dependence of the stabilized frequency step $\bar{\Delta}_{\rm step}(\Delta_{\rm sh})$, which has the universal form: $\bar{\Delta}_{\rm step}=\Delta_{\rm sh}$ for arbitrary values of $\Omega_0$, $\tau^{}_1$ and $\tau^{}_2$.} \label{Fig4}
\end{figure}

As an alternative to the previous method with additional varied phase $\phi^{c}$ during the second pulse \cite{Sanner_2017}, let us describe another variant of GABRS, where the concomitant parameter $\xi$ is equal to the varied additional frequency step $\Delta_{\rm step}$ during both Ramsey pulses $\tau^{}_1$ and $\tau^{}_2$ (Fig.~\ref{Fig4}(a)). This frequency-step technique was proposed in Refs.~\cite{tai09,yudin2010}. Excluding the explicit phase jumps $\alpha_{\pm}^{}$, the frequency step $\Delta_{\rm step}$ can be formally described by a phase function $\varphi (t)$ in Eq.~(\ref{E}) with non-zero time derivative $d\varphi (t)/dt=\Delta_{\rm step}$ during the pulses ($\tau^{}_1$ and $\tau^{}_2$) and with zero time derivative $d\varphi (t)/dt=0$ during the dark time $T$, and phase continuity is maintained throughout. In this case, we always have $\bar{\delta}_{\rm clock}=0$, and the stabilized frequency step $\bar{\Delta}_{\rm step}$ is determined as the solution of Eq.~(\ref{eq_alpha}), which has universal form: $\bar{\Delta}_{\rm step}=\Delta_{\rm sh}$ for arbitrary values of $\Omega_0$, $\tau^{}_1$ and $\tau^{}_2$ (Fig.~\ref{Fig4}(b)). This universal dependence can be slightly deformed only due to some non-idealities of the interrogation scheme (e.g., if $\alpha^{}_{+}\neq -\alpha^{}_{-}$).

\begin{figure}[t]
\centerline{\scalebox{0.5}{\includegraphics{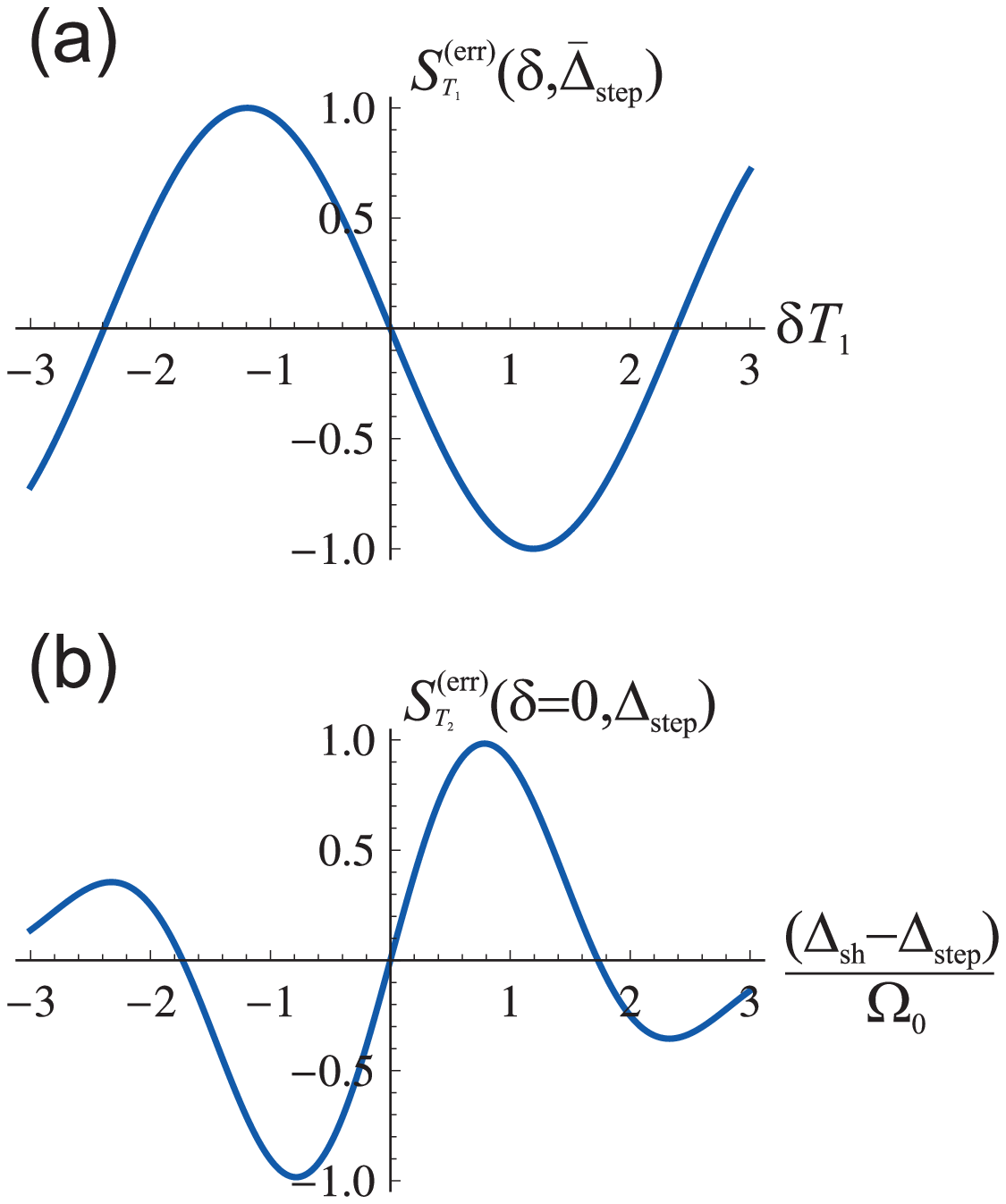}}}\caption{
Error signals under $\Omega_0 T_1=2\pi$, $\Omega_0 \tau^{}=\pi/2$ ($\tau^{}_1=\tau^{}_2=\tau^{}$), and for arbitrary field-induced shifts of the clock transition during Ramsey pulses, $\Delta_{\rm sh}$:\\
(a) Error signal $S^{\rm (err)}_{T_1}(\delta,\Delta_{\rm step}$=$\bar{\Delta}_{\rm step})$; (b) Error signal $S^{\rm (err)}_{T_2}(\delta=0,\Delta_{\rm step})$.} \label{Fig5}
\end{figure}

The error signals $S^{\rm (err)}_{T_1}(\delta,\Delta_{\rm step}$=$\bar{\Delta}_{\rm step})$ and $S^{\rm (err)}_{T_2}(\delta$=$0,\Delta_{\rm step})$ have universal antisymmetrical forms for different values of $\Delta_{\rm sh}$ (Fig.~\ref{Fig5}). Note that this antisymmetry does not depend on the Rabi frequency $\Omega_0$. Thus, we believe that this variant of GABRS is more optimal and robust than the approach used in Ref.~\cite{Sanner_2017} where an additional phase $\phi^{c}$ was varied during the second pulse. In fact, in the case of $|\Delta_{\rm sh}/\Omega_0|\gg 1$ it is already necessary to use the frequency-step technique of Refs.~\cite{tai09,yudin2010} to compensate for the very large actual shift $\Delta_{\rm sh}$. This frequency-step technique was also used in experiments in Ref.~\cite{Sanner_2017} with the Yb$^+$ ion, because for the octupole clock transition the condition $|\Delta_{\rm sh}/\Omega_0|\gg 1$ is practically always true. But in this case, the use of an additional varied phase $\phi^{c}$ in Ref.~\cite{Sanner_2017} seems to be an excessive technical complication, because we can directly use the frequency-step technique ($\Delta_{\rm step}$) in GABRS without any additional manipulations.

Note that this variant of GABRS is suitable also for CPT atomic clocks, when we can use as concomitant parameter $\xi$ the varied frequency step $\Delta_{\rm step}$ during both Ramsey pulses in CPT-Ramsey spectroscopy.

\subsection{Auto-balanced Ramsey spectroscopy with varied pulse duration}

\begin{figure}[t]
\centerline{\scalebox{0.5}{\includegraphics{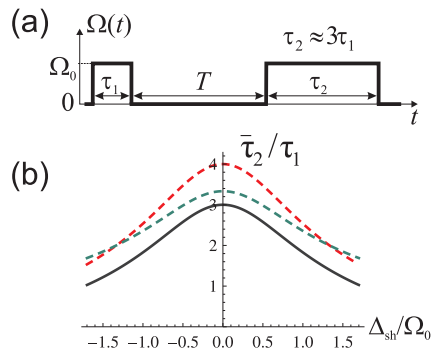}}}\caption{
(a) Schematic illustration of a Ramsey interrogation scheme for GABRS, where the concomitant parameter $\xi$ is equal to the duration of the second pulse $\tau^{}_2$.\\
(b) The dependence of stabilized pulse duration $\bar{\tau}^{}_2(\Delta_{\rm sh})$ for different pulse area: $\Omega_0\tau^{}_1=\pi/2$ (black solid line), $\Omega_0\tau^{}_1=0.8\times\pi/2$ (red dashed line), $\Omega_0\tau^{}_1=1.2\times\pi/2$ (green dashed line).} \label{Fig6}
\end{figure}

\begin{figure}[t]
\centerline{\scalebox{0.5}{\includegraphics{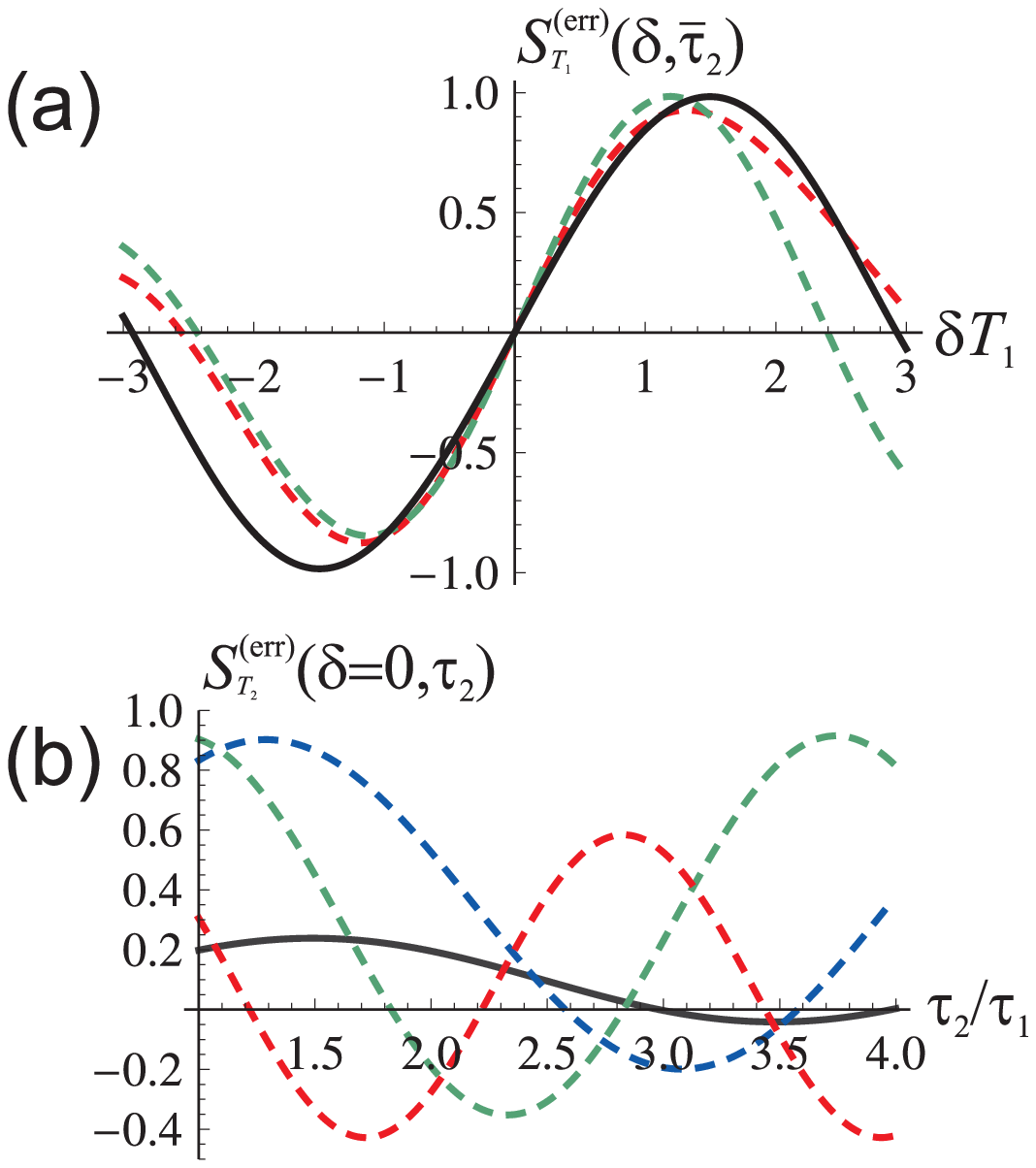}}}\caption{
Error signals under $\Omega_0 T_1=2\pi$, $\Omega_0 \tau^{}_1=\pi/2$, and for different field-induced shifts of the clock transition during Ramsey pulses, $\Delta_{\rm sh}$:\\
(a) Error signal $S^{\rm (err)}_{T_1}(\delta,\tau^{}_2=\bar{\tau}^{}_2)$: $\Delta_{\rm sh}/\Omega_0=0$ (black solid line); $\Delta_{\rm sh}/\Omega_0=0.5$ (red dashed line); $\Delta_{\rm sh}/\Omega_0=1.0$ (green dashed line).\\
(b) Error signal $S^{\rm (err)}_{T_2}(\delta=0,\tau^{}_2)$: $\Delta_{\rm sh}/\Omega_0=0.1$ (black solid line); $\Delta_{\rm sh}/\Omega_0=0.5$ (blue dashed line); $\Delta_{\rm sh}/\Omega_0=1.0$ (green dashed line); $\Delta_{\rm sh}/\Omega_0=1.5$ (red dashed line). The case of $\Delta_{\rm sh}/\Omega_0=0$ is not included here, because the error signal $S^{\rm (err)}_{T_2}(\delta=0,\tau^{}_2)$ approaches zero for any $\tau^{}_2$ in this limit, $|\Delta_{\rm sh}/\Omega_0|\rightarrow 0$.} \label{Fig7}
\end{figure}
For generality, let us describe a variant of GABRS, where the concomitant parameter $\xi$ is equal to the varied duration of the second (as an example) Ramsey pulse $\tau^{}_2$, while the pulse duration of the first Ramsey pulse $\tau^{}_1$ is fixed. For this method we will use the Ramsey sequence, which was considered in Ref.~\cite{yudin2010}, where $\tau^{}_2\approx 3\tau^{}_1$ (Fig.~\ref{Fig6}(a)). In this case, we always have $\bar{\delta}_{\rm clock}=0$, and the stabilized pulse duration $\bar{\tau}^{}_2$ is determined as the solution of Eq.~(\ref{eq_alpha}). In the presence of the field-induced shift of the clock transition $\Delta_{\rm sh}$ during the Ramsey pulses, the duration $\bar{\tau}^{}_2$ is a function $\bar{\tau}^{}_2(\Delta_{\rm sh})$ on the value $\Delta_{\rm sh}$. These dependencies are presented in Fig.~\ref{Fig6}(b) for different pulse areas of the first Ramsey pulse, $\Omega_0 \tau^{}_1$. The error signals $S^{\rm (err)}_{T_1}(\delta,\tau^{}_2=\bar{\tau}^{}_2)$ and $S^{\rm (err)}_{T_2}(\delta=0,\tau^{}_2)$ for different values $\Delta_{\rm sh}$ are presented in Fig.~\ref{Fig7}.

Note that this variant of GABRS is not valid for CPT atomic clocks.

\section{Conclusion}
We have developed a method and theoretical basis of generalized auto-balanced Ramsey spectroscopy (GABRS), which allows for the elimination of probe-field-induced shifts in atomic clocks.
This universal two-loop method requires the use of a concomitant parameter $\xi$ in addition to the clock frequency $\omega$, which is related to the first and/or second Ramsey pulses $\tau^{}_1$ and $\tau^{}_2$ through the use of interleaved Ramsey sequences with two different dark times $T_1$ and $T_2$. A correlated stabilisation of both variable parameters can be achieved. It was analytically shown that the  GABRS method always leads to zero field-induced shift of the stabilized frequency $\omega$ in an atomic clock, $\bar{\delta}_{\rm clock}=0$, independent of relaxation processes (including decoherence) and different imperfections in the interrogation procedure. Such robustness is a direct consequence of the phase-jump technique used to build an error signal in Ramsey spectroscopy. We have considered several variants of GABRS with the use of different concomitant parameters $\xi$. It was found that the most optimal and universal variant is based on the frequency-step technique, where the concomitant parameter $\xi$ is a varied additional frequency step $\Delta_{\rm step}$ during both Ramsey pulses $\tau^{}_1$ and $\tau^{}_2$. In this case, universal anti-symmetrical error signals are generated, which result in vanishing frequency shift even at finite modulation amplitude. Some variants of GABRS can also be applied to CPT atomic clocks using CPT-Ramsey spectroscopy of the two-photon dark resonance. Moreover, GABRS is valid for open systems, and therefore, this technique can be exploited with more complex schemes, such as molecules for high-resolution molecular spectroscopy \cite{Eibach_2011,George_2011}.
It is also possible that more complicated Ramsey pulse sequences, for example hyper-Ramsey sequences \cite{yudin2010}, could also take advantage of the generalized auto-balance techniques described here. Note that the experimental results in Ref.~\cite{Sanner_2017} can be considered as a first confirmation of the GABRS theory developed in our paper.

\begin{acknowledgments}
We thank Ch. Sanner, Ch. Tamm, E. Peik, N. Huntemann, K. Beloy and C. Oates for useful discussions and comments. This work was supported by the Russian
Scientific Foundation (No. 16-12-10147). This work is also funded by NIST, an agency of the U.S. government, and is not subject to copyright.
\end{acknowledgments}

\end{document}